\documentclass[journal=apchd5,manuscript=article,layout=twocolumn]{achemso}

\usepackage{amsmath}
\usepackage{amssymb}
\usepackage{amsfonts}
\usepackage{mathtools}
\usepackage{physics}
\usepackage{cuted}
\usepackage{xcolor}

\author{Vashist G. Ramesh}
\author{Joris Busink}
\author{Ren\'e E. R. Moesbergen}
\author{Kevin J. H. Peters}
\author{Philip J. Ackermans}
\author{Said K. R. Rodriguez}
\affiliation{Center for Nanophotonics, AMOLF, Science Park 104, 1098 XG Amsterdam, The Netherlands}
\email{s.rodriguez@amolf.nl}

\title[]
{Stochastic Thermodynamics of a Linear Optical Cavity Driven On Resonance}

\abbreviations{CFT,JE,PDF,FT,OLE}

\begin{document}
	
	
	
	
	
	
	
\begin{abstract}
We present a complete framework of stochastic thermodynamics for a single-mode linear optical cavity driven on resonance. We first show that the steady-state intra-cavity field follows the equilibrium Boltzmann distribution. The effective temperature is given by the noise variance, and the equilibration rate is the dissipation rate.  Next we derive expressions for  internal energy, work, heat, and free energy of light in a cavity, and formulate the first and second laws of thermodynamics for this system.  We then analyze fluctuations in work and heat, and show that  they obey universal statistical relations known as fluctuation theorems. Finite time corrections to the fluctuation theorems are also discussed. Additionally, we show that work fluctuations obey the Crook’s Fluctuation theorem which is a paradigm for understanding emergent phenomena and estimating free energy differences. The significance of our results is two-fold. On one hand, our work positions optical cavities as a unique platform for fundamental studies of stochastic thermodynamics. On the other hand, our work paves the way for improving the energy efficiency and information processing capabilities of laser-driven optical resonators using a thermodynamics based prescription. 	\end{abstract}
	
\section{Introduction} \label{S1}
Science usually precedes technology, but thermodynamics is an exception. Steam engines worked before thermodynamic laws were discovered. Actually, thermodynamics  emerged from the desire to increase engine efficiencies. Eventually, the formulation and experimental validation of thermodynamic laws yielded more than better engines. Thermodynamics earned the timeless authority to determine which processes are possible, and to discard those ideas that do not abide by its principles. To date, technologies are conceived and optimized based on thermodynamics. This work is motivated  by the conviction that many nanophotonic devices are now at a stage comparable to that of early steam engines. These devices can be made and characterized on astonishingly small scales thanks to nanotechnology, but a framework to increase their energy efficiency (with minimum sacrifice in speed and precision) in the inevitable presence of noise is lacking. The standard optics framework, based on deterministic Maxwell’s equations, cannot solve this issue because it neglects noise. However, a second chapter in the history of thermodynamics points at a solution.
	
\begin{figure}[t]
		\includegraphics[keepaspectratio]{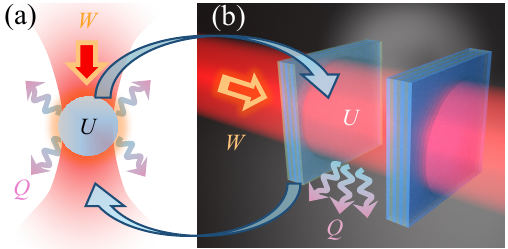}
		\caption{(a) A laser-trapped particle as widely studied in stochastic thermodynamics. $U$ is the internal energy, $W$ the work, and $Q$ the heat dissipated to the environment. (b) A single-mode Fabry-Perót cavity, as studied in this manuscript. Notice how the roles of light and matter are reversed. In (a) the trapping potential is light, and the system of interest is made of matter.  In contrast, in (b) the trapping potential is made of matter, while the system of interest is light. }\label{fig1}
	\end{figure}

Over the past 25 years, Stochastic Thermodynamics (ST) emerged as a comprehensive framework for describing small energy-harvesting and information-processing systems in contact with heat or chemical reservoirs~\cite{Jarzynski11, Seifert12, Parrondo15, Ciliberto17}. Consider, for example, a laser-trapped colloidal particle as shown in Fig.~\ref{fig1}(a). This system works as a micron-scale heat engine~\cite{Blickle12, Martinez16}, with laser and particle respectively replacing the piston and working gas. The first law of thermodynamics, $\delta U = W - Q$ relates the system’s change in internal energy $\delta U$ to the work $W$ done on the system (usually negative for an engine) and the dissipated heat $Q$.  ST is concerned with fluctuations in these and other thermodynamic quantities, which are prominent in small systems. By accounting for these fluctuations, ST places fundamental limits on energy and information processing capabilities of materials.  ST advances the types of ideas needed to address current and emerging challenges in nanophotonics, many of which are related to stochastic effects. However, relations between optical and thermodynamic quantities need to be established.

Since photons typically do not reach thermal equilibrium (except under special conditions~\cite{Klaers10T}), thermodynamics is rarely used to describe states of light and their transformations. Recently, however, thermodynamics has been increasingly used to understand how material properties limit or enable optical functionalities~\cite{LeThomas18,Miller18, Mann19, Hamerly20, Manjavacas21, Fan22, Monticone22}. In addition,  thermodynamic concepts have enabled the discovery and engineering of fascinating phenomena in multi-mode optical systems~\cite{Wu2019,Muniz23,Pyrialakos22}. Some aspects of stochastic quantum thermodynamics have been theoretically explored in optical resonators~\cite{Kewming22}.  However,  a classical framework of stochastic thermodynamics  has never been presented for a single-mode linear optical cavity. Filling this important knowledge gap is the goal of this manuscript.

Here we present a complete  stochastic thermodynamic framework for a coherently and resonantly driven linear optical cavity. This manuscript is organized as follows. In Section~\ref{S2} we introduce the model for our system, and derive the scalar potentials confining light.  In Section~\ref{S3}  we show effective equilibrium behavior of light in a resonantly-driven cavity. The steady-state intra-cavity field is shown to follow the equilibrium Boltzmann distribution, and an expression for the partition function is presented. In Section~\ref{S4} we formulate the first and second laws of thermodynamics for our system. In Section~\ref{S5} we analyze the averaged work and heat generated when modulating the laser amplitude. We elucidate how non-equilibrium behavior emerges when the modulation time is commensurate with the dissipation time. In Section~\ref{S6} we analyze  work and heat fluctuations, and show that they obey universal statistical relations known as Fluctuation Theorems (FTs). We furthermore show that light in the cavity satisfies Crook’s Fluctuation theorem (CFT), enabling the estimation of free energy differences based on non-equilibrium work measurements.  Finally, in Section~\ref{S7} we summarize our results and discuss perspectives they offer.

\section{The model} \label{S2}
We consider a single-mode coherently-driven linear optical resonator. We envision a laser-driven plano-concave Fabry-Per\'ot cavity for concreteness, as illustrated in Fig.~\ref{fig1}(b). However, our model equally describes any coherently-driven resonator provided that two conditions are fulfilled. First, one mode needs to be sufficiently well isolated, spectrally and spatially, from all other modes. Second, the laser intensity  needs to be sufficiently low for linear response to hold. These conditions can be fulfilled in open cavities~\cite{Mader22}, whispering-gallery-mode~\cite{Vollmer24}, photonic crystal~\cite{Perrier20}, or plasmonic~\cite{Li24} resonators, for example.

In a frame rotating at the laser frequency $\omega$, the field $\alpha$ in the cavity obeys the following equation of motion:
	\begin{equation} \label{eq1}
		i\dot{\alpha} = \left(-\Delta- i\frac{\Gamma}{2}\right)\alpha + i\sqrt{\kappa_L}A + D\zeta(t).
	\end{equation}
\noindent  $\Delta=\omega-\omega_0$ is the detuning between $\omega$ and the cavity resonance frequency $\omega_0$. $\Gamma = \gamma_a +\kappa_L+\kappa_R$ is the total loss rate, comprising the absorption rate $\gamma_a$ and input-output rates $\kappa_{L,R}$ through the left and right mirrors. $A$ is the laser amplitude, which we assume to be real. $D \zeta(t)$ is a stochastic force comprising a complex-valued Gaussian process $ \zeta(t)= \zeta_R(t) + i  \zeta_I(t)$ with mean $\langle\zeta_{R,I}(t)\rangle=0$ and correlation $\langle\zeta_{R}(t)\zeta_{I}(t')\rangle = \delta_{RI} \delta(t-t')$. The constant $D$ is the standard deviation of the stochastic force.

Our model accounts for two sources of noise in every coherently-driven resonator. One of them is the noise of the incident laser. The other is the dissipative interaction of the cavity with its environment. According to the fluctuation-dissipation relation, that interaction results in fluctuations of the intra-cavity field. We can use a single pair of stochastic terms $\zeta_{R,I}$ to effectively account for both noise sources under the reasonable assumption that they are additive and Gaussian. Reference~\citenum{Ramesh24} presents one of many examples in the literature of an experimental system described by our model.

To analyze the deterministic force acting on $\alpha$, we decompose Equation~\ref{eq1} into real and imaginary parts. Setting $\alpha=\alpha_R+i\alpha_I$, we get
	\begin{align}  \label{eq2}
		\begin{pmatrix}\dot{\alpha_R} \\ \dot{\alpha_I}\end{pmatrix}=
		\underbrace{\begin{pmatrix}-\frac{\Gamma}{2} & -\Delta \\ \Delta &-\frac{\Gamma}{2}
			\end{pmatrix}
			\begin{pmatrix}\alpha_R \\ \alpha_I \end{pmatrix}  + \begin{pmatrix} \sqrt{\kappa_L}A\\ 0 \end{pmatrix}}_{\textbf{F}/\Gamma} + D\vec{\zeta}.
	\end{align}
\noindent Equation~\ref{eq2} is a two-dimensional overdamped Langevin equation (OLE).  The underbraced term  contains the deterministic force $\textbf{F}$, divided by $\Gamma$ to recover the normal form of the OLE.

$\textbf{F}$ is fundamentally different when $\Delta \neq 0$ or $\Delta=0 $. When $\Delta \neq 0$, $\textbf{F}$ contains a conservative and a non-conservative part~\cite{Peters23}. A conservative force is one that can be derived from a scalar potential $U$, i.e., $\textbf{F}_c=-\nabla U$. A non-conservative force is equal to the curl of a vector potential: $\textbf{F}_{n}=\nabla \cross \mathbf{A}$. This manuscript focuses entirely on the case $\Delta=0$, wherein the cavity is driven exactly on resonance and $\textbf{F}=\textbf{F}_c$. Only in this case, an equilibrium steady-state can be expected~\cite{Chernyak06, Peters23}.

When $\Delta=0 $, Eq.~\ref{eq2} decouples into a pair of independent one-dimensional OLEs:
	
 \begin{equation} \label{OLE1D}
		\Gamma\dot\alpha_{R,I} = - \frac{\partial U_{R,I}}{\partial \alpha_{R,I}} +  \Gamma D\zeta_{R,I}(t).
	\end{equation}
The potential energies $U_{R,I}$ are obtained by integrating the deterministic forces in Eq.~\ref{eq2}:
	\begin{subequations}\label{V}
		\begin{align}
			U_{R} &= \frac{\Gamma^2}{4}\alpha_{R}^2 - \Gamma\sqrt{\kappa_L}A\alpha_{R}, \label{VR}\\
			U_{I} &= \frac{\Gamma^2}{4}\alpha_{I}^2.	\label{VI}
		\end{align}
	\end{subequations}
\noindent The potentials are harmonic, as expected for a linear cavity. Their only difference is that the minimum of $U_R$ is shifted from zero by the incident laser amplitude.

\section{Effective equilibrium} \label{S3}
\begin{figure}[t]
		\includegraphics{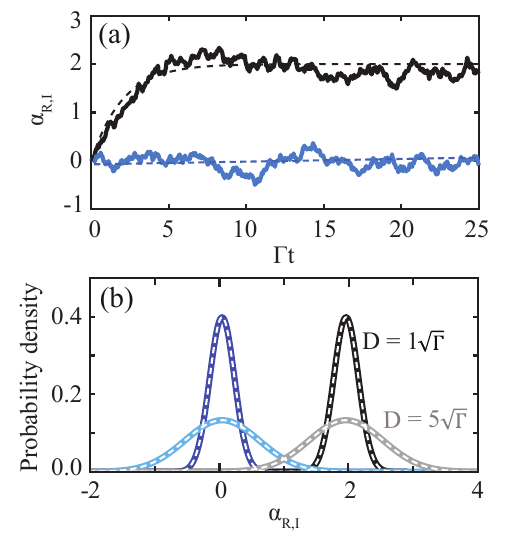}
		\caption{(a) Sample trajectories of the real and imaginary parts of the field, $\alpha_R$ and $\alpha_I$ in black and blue, respectively. Solid curves are numerical simulations of Eq.~\ref{OLE1D}. Dashed curves are theoretical predictions from Eq.~\ref{solutions}. (b) Probability density functions (PDFs) of $\alpha_R$  and $\alpha_I$ in black and blue, respectively. Dark and light shades correspond to two different noise standard deviations $D$. White dashed curves are theoretical distributions using Eq.~\ref{Bol} and Eq.~\ref{Z}. Model parameters are $\Delta = 0$, $A = \sqrt{2\Gamma}$, $\Gamma=1$, $\kappa_L=\Gamma/2$.}\label{fig2}
	\end{figure}

In Fig.~\ref{fig2} we present numerical solutions to Eq.~\ref{OLE1D}, obtained using the xSPDE \textsc{Matlab} toolbox~\cite{xspde}.  Figure~\ref{fig2}(a) shows sample trajectories of $\alpha_R$ and $\alpha_I$ as black and blue curves, respectively. Based on $20000$ of such trajectories, we constructed probability density functions (PDFs) of $\alpha_R$ and $\alpha_I$; these are shown as black and blue curves in Fig.~\ref{fig2}(b), respectively. For both $\alpha_R$ and $\alpha_I$, we consider two different standard deviations of the noise $D$. In the following, we explain how Fig.~\ref{fig2} displays effective thermodynamic equilibrium behavior. This behavior is present both at the level of the individual trajectories and of the PDFs.

Figure~\ref{fig2}(a) shows $\alpha_{R}$ rising to a steady state and fluctuating thereafter. Meanwhile, $\alpha_{I}$ fluctuates around 0 (its steady state value) all the time. We can calculate the deterministic evolution of $\alpha_{R.I}$ by solving Eq.~\ref{OLE1D}, with $D=0$, analytically:
\begin{subequations}\label{solutions}
		\begin{align}
    \left<\alpha_R(t)\right> &= \frac{2\sqrt{\kappa_L}A}{\Gamma} \left(1 - \exp\left(-\frac{\Gamma t}{2}\right)\right)\\
    \left<\alpha_I(t)\right> &= 0.
    \label{eq:alphaConstant}
\end{align}
\end{subequations}

\noindent The above solutions are plotted as dashed curves on top of the numerical results in Fig.~\ref{fig2}(a). Notice that
$\Gamma^{-1}$ is the characteristic time in which $\alpha_R$ reaches its steady state. Since the steady-state distribution is the equilibrium Boltzmann distribution (shown next), $\Gamma^{-1}$ is also the equilibration time of the fields.

The PDF of a gas in thermal equilibrium, confined in a scalar potential $U$, is the well-known equilibrium Boltzmann distribution $Z^{-1}\exp{-\beta U}$. $Z$ is the partition function, and $\beta = 1/k_B T$ with $k_B$ Boltzmann's constant and $T$ the temperature. Following Peters \textit{et al}.~\cite{Peters23}, we can relate the noise variance $D^2$ to the effective temperature of the light field via the fluctuation dissipation relation $\Gamma D^2 = 2k_BT$. Using this relation and the scalar potentials $U_{R,I}$ in Equation~\ref{V}, we arrive to the following expression for the Boltzmann distribution of the intra-cavity field:

\begin{equation} \label{Bol}
		\mathcal{P}(\alpha_{R,I}) = Z^{-1}_{R,I} \exp\left(-\frac{2 U_{R,I}}{\Gamma D^2} \right)
	\end{equation}
\noindent

The partition functions can be obtained by imposing the normalization condition on Eq.~\ref{Bol}, i.e., $\int_{-\infty}^{\infty}\mathcal{P}(\alpha_{R,I})d\alpha_{R,I}=1$. Doing this for $\alpha_R$ we obtain:
\begin{equation} \label{Z}
		Z_R = \sqrt{\frac{ 2\pi D^2}{\Gamma}}\exp\left(\frac{2\kappa_LA^2}{\Gamma D^2}\right).
\end{equation}
\noindent Notice that both the equilibrium Boltzmann distribution and the partition function are written in terms of the experimentally accessible standard deviation of the noise $D$ and dissipation (cavity linewidth) $\Gamma$, instead of $\beta$ which cannot be directly measured.  In the next section we will use the expression for $Z_R$ to formulate the second law of thermodynamics.

The white dashed lines in Fig.~\ref{fig2}(b) were calculated using  Eqs.~\ref{Bol} and~\ref{Z}. Their excellent agreement with the numerically calculated distributions demonstrates that light confined in a linear optical  resonator displays effective thermal equilibrium behavior: the steady-state distribution is the equilibrium Boltzmann distribution. The effective temperature is related to the noise according to the aforementioned fluctuation-dissipation relation.  For a detailed discussion about the meaning of the effective temperature $T$, we refer to Ref.~\citenum{Peters23}.

\section{First  and second law of thermodynamics} \label{S4}

In this section we formulate the first and second law of thermodynamics for our resonantly-driven linear optical cavity. Starting from Eq.~\ref{OLE1D}, we apply the approach of Sekimoto~\cite{Sekimoto98} to derive expressions for the internal energy, work, and heat. For brevity we  only consider Eq.~\ref{OLE1D} with $U_R$. However, our results can be trivially extended to the case of Eq.~\ref{OLE1D} with $U_I$ by setting $A=0$.

We first multiply both sides of Eq.~\ref{OLE1D} with an infinitesimal field change $d \alpha_R$:
\begin{equation} \label{eq13}
		\Gamma\dot\alpha_R d \alpha_R  = -\frac{\partial U_R}{\partial \alpha_R} d \alpha_R + \Gamma D\zeta_R d \alpha_R.
\end{equation}
\noindent Next, we use the expression for the total differential of $U_R=U_R(\alpha_R , A)$,  which according to the chain rule is: $d U_R = \frac{\partial U_R}{\partial \alpha_R}d\alpha_R +  \frac{\partial U_R}{\partial A}dA$. Using this expression for $d U_R$, and the relation  $d \alpha_R = \dot\alpha_R d t$, we obtain from Eq.~\ref{eq13}:
\begin{equation} \label{eq14}
	\Gamma\dot\alpha_R^2 d t  = - d U_R + \frac{\partial U_R}{\partial A}dA + \Gamma D\zeta_R \dot\alpha_R d t.
\end{equation}	
\noindent We now substitute $d A = \dot Ad t$ and rearrange terms to get:
\begin{equation} \label{1stlawd}
	d U_R =  \underbrace{- \Gamma\sqrt{\kappa_L}\alpha_R\dot{A}dt}_{+dW} + \underbrace{\Gamma (D\zeta_R \dot\alpha_R d t - \dot\alpha_R^2 d t)}_{-dQ}.
\end{equation}	

\noindent Using the expression for $U_R$  in Eq~\ref{VR} and integrating all terms in time (across an arbitrary trajectory from $t=0$ to $t=s$) we get:
\begin{subequations} \label{UWQ}
		\begin{align}
			\delta U_R &= \left[\frac{\Gamma^2}{4}\alpha_R(t)^2 -\Gamma\sqrt{\kappa_L}A(t)\alpha_R(t)\right]_{0}^{s}, \label{UWQa}\\
			W &= -\int_{0}^{s}\Gamma\sqrt{\kappa_L}\alpha_R(t)\dot{A}(t) d t, \label{UWQb}\\
			Q &= \int_{0}^{s}\Gamma\left(\dot{\alpha_R}(t)^2 - D\zeta_R(t)\dot{\alpha_R}(t)\right)  d t.\label{UWQc}
		\end{align}
\end{subequations}

\noindent Using the above expressions, we can now formulate the first law of thermodynamics for a resonantly-driven stochastic linear optical cavity:
\begin{equation} \label{1stlaw}
	\delta U_R =  W- Q.
\end{equation}

$\delta U_R$ is the net change in internal energy of the cavity over the trajectory. To recognize this, consider that $U_R$ is proportional $\alpha_R^2$. For $\Delta=0$, as considered throughout this manuscript, $\alpha_R^2$ is the number of intra-cavity photons which is also the energy stored in the cavity.

$W$ is the work done by the laser field on the intra-cavity field. To recognize this, notice that the integrand in Eq.~\ref{UWQb} contains the product of $\alpha_R$ and $\Gamma \sqrt{\kappa_L} \dot{A}$, which is the time derivative of the force due to the laser.  Interpreting $\alpha_R$ as a displacement and integrating Eq.~\ref{UWQb} results in a force times displacement, i.e., work. $W>0$ means work is done on the intra-cavity field. The form of the work in Eq.~\ref{UWQb}, introduced by Jarzynski~\cite{Jarzynski97}, is in general different from the ``classical work'' as known in the statistical physics literature~\cite{Hummer01,Douarche05}. The latter is defined as the time-integral of a force times a velocity. In particular, for $\dot{A}=0$ (constant laser amplitude) the so-called Jarzynski work is zero but  the classical work is not. The two works are only equivalent for periodic driving $A(t)=A(t+\tau)$, with $\tau$ the period.

The heat $Q$ in Eq.~\ref{UWQc} quantifies the transport of energy from the cavity to its environment. It contains two terms. The first term is the time-integrated dissipated power,  given by the velocity squared as expected for a harmonic oscillator. The second term contains the product of the  stochastic force $D\zeta_R(t)$ and the velocity $\dot{\alpha_R}$, integrated over time. This is precisely the classical work done by the environment on the intra-cavity field. Thus, the net heat transfer is given by the difference between the dissipated energy to the environment and the work done by the environment on the system.

We now proceed to formulate the second law of thermodynamics for our system. The second law states
\begin{equation}\label{2ndlaw}
	\langle W\rangle\geq\delta F,
\end{equation}
\noindent with $\langle W\rangle$ the average work and $\delta F$ the free energy difference between initial and final states. The lower bound $\langle W\rangle=\delta F$ is only attained by a reversible process. Notice that, unlike the first law, the second law does not hold at the level of individual trajectories. Actually, in the early days of stochastic thermodynamics, individual trajectories with $W < \delta F$ were occasionally called ``transient violations of the second law''~\cite{Wang02, Ritort04, Wang05}. Those were not really violations of the second law of course, which applies only on average~\cite{Seifert12}.

$W$ for our resonantly-driven stochastic linear optical cavity was already defined in Eq.~\ref{UWQb}. Hence, to formulate the second law for our system we only need to define the free energy $F$. We can easily get this from the relation
\begin{equation} \label{F}
		\beta F = -\ln{Z_R},
	\end{equation}
\noindent with $Z_R$ as defined in Eq.~\ref{Z}. Like the internal energy, $F$ has units of $k_BT$. It does not depend explicitly on time or $\alpha$. It is therefore an intensive quantity that does not fluctuate. In the next section we show, through numerical simulations of our cavity system, that the second law is indeed always respected. However, in a  finite-time trajectory there is a nonzero probability for $W < \delta F$. That probability is quantified by a fluctuation theorem.

We close this section by noting that all our definitions (see, particularly, Eqs.~\ref{UWQ} and~\ref{F}) possess consistent thermodynamic interpretations, but do not possess true units of energy. Notice in our fluctuation-dissipation relation $\Gamma D^2 = 2k_BT$ that $k_BT$ has units $\Gamma^2$. The extra factor of $\Gamma$ is due to the fact that while Eq.~\ref{eq1} has the form of the OLE, the dissipation $\Gamma$ of our optical cavity is in the right hand side of the equation. In contrast, the standard OLE reads $\gamma \dot{x} = \mathbf{F} + \xi(t)$, with  $\gamma$ the dissipation, $\mathbf{F}$ the deterministic force, and $\xi(t)$ the stochastic force. Because of this difference, our thermodynamic description is only effective. The effective character of our description is also evident in our interpretation of $\alpha$ as a displacement (with units of meters) when defining work and heat. In our single-mode cavity model, however, $\alpha$ is dimensionless. Despite these differences which make the interpretation of our results subtle,  our framework is nonetheless complete and self-consistent in the sense that all fundamental relations between (effective) thermodynamic quantities hold.

\section{Averaged thermodynamic quantities under periodic driving} \label{S5}

In this section we discuss averaged
thermodynamic quantities under time-harmonic driving.  Protocols of this kind have been widely studied in stochastic thermodynamics~\cite{Blickle06,Schuler05,Ciliberto17,Joubaud07,Douarche06}. They have the benefit of generality --- any protocol  can be decomposed into sine and cosine modes via a Fourier transform.
	
	\begin{figure*}[t]
		\includegraphics[]{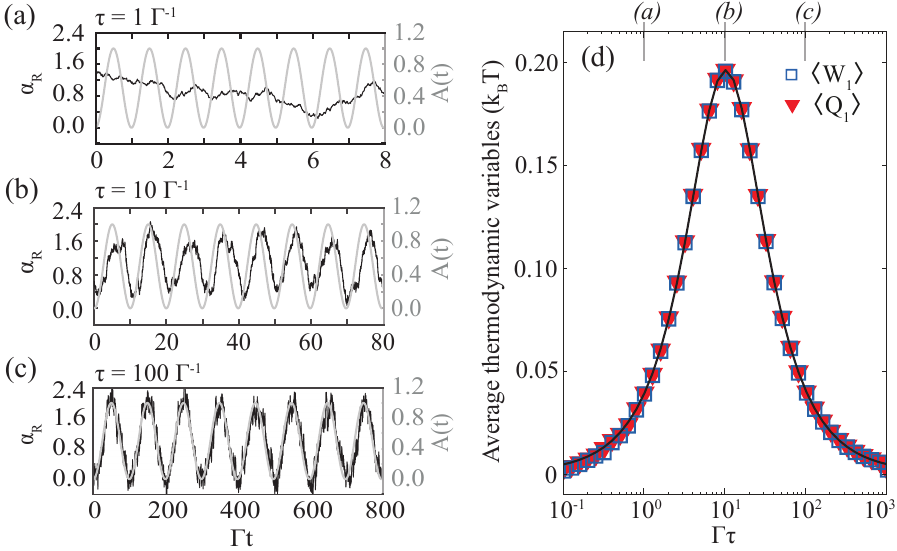}
		\caption{(a)-(c) Gray and black curves are the laser amplitude $A$ and intra-cavity field $\alpha_R$, respectively. The modulation period $\tau$ is indicated in each panel. (d) Average work $\langle W_1\rangle$ and heat $\langle Q_1\rangle$ done in one modulation period of duration $\tau$. The black curve is the theoretical prediction in Eq.~\ref{avg}. Model parameters:  $D = \sqrt{\Gamma/2}$, $\kappa_L=\Gamma/2$, $A=A_0(1+\cos(2\pi t/\tau))$, $A_0=\sqrt{\Gamma}/2$.}\label{fig3}
	\end{figure*}

Figures~\ref{fig3}(a)-(c) show trajectories of $\alpha_R$ for one realization of the noise and  three distinct $\tau$. $A$ and $\alpha_R$ are plotted as gray and black curves, respectively. Figure~\ref{fig3}(a) was obtained for $\tau= \Gamma^{-1}$, which results in highly non-adiabatic dynamics. $\alpha_R$ cannot follow the driving laser because its instantaneous rate of change is too large compared to the equilibration time $\Gamma^{-1}$. Next, Fig.~\ref{fig3}(b) was obtained for $\tau=10\Gamma^{-1}$. Here the trajectory of $\alpha_R$ resembles the driving protocol, but there is a delay which results in hysteresis. This hysteresis is due to the fact that $\alpha_R$ does not fully equilibrate at any point in time during the protocol. While the period exceeds the equilibration time, the amplitude of the modulation is so large that the system is constantly driven out of equilibrium. Finally, Fig.~\ref{fig3}(c) was obtained for $\tau= 100\Gamma^{-1}$. Here the driving is adiabatic and $\alpha_R$ closely follows the laser. The system constantly remains in a state of local equilibrium. Overall, Figs.~\ref{fig3}(a)-(c) illustrate how non-equilibrium behavior emerges when the intra-cavity field changes within a time that is commensurate with, or shorter than,  $\Gamma^{-1}$.
	
In Fig.~\ref{fig3}(d) we analyze the average work and heat (as defined in Eq.~\ref{UWQ}) produced in one period of the modulation in $A$. Averages are done over $2000$ modulation periods, and we plot the results as a function of $\tau$. Notice that the average work and heat are always equal to each other. This is a consequence of the first law combined with a net zero change in average internal energy. The average internal energy does not change because initial and final states in our periodic protocol are the same. Further, notice in Fig.~\ref{fig3}(d) that $\langle W_1 \rangle\geq0= \delta F$, as expected from the second law. Moreover, the lower bound $\langle W_1 \rangle =\delta F=0$ is attained in the adiabatic limit $\tau \to \infty$, wherein the system remains in equilibrium and the dynamics are reversible.

On top of the numerical simulations in  Fig.~\ref{fig3}(d), we plot the theoretical prediction for the average work and heat as a black curve. This was obtained by setting $A(t)=A_0(1+\cos(2\pi t/\tau))$ and $D=0$ in Eqs.~\ref{OLE1D} and~\ref{UWQ}, which results in the expression
    \begin{equation} \label{avg}
		\langle W_n\rangle = n\frac{2\pi^2\kappa_LA_0^2\Gamma\tau}{16\pi^2+\Gamma^2\tau^2}=\langle Q_n\rangle.
	\end{equation}
The subscript $n$ is the number of modulation periods integrated over, which is equal to one for the results in Fig.~\ref{fig3}(d). The theoretical prediction is in excellent agreement with the simulations.

Figure~\ref{fig3}(d)  shows that $\langle W_1\rangle$ and $\langle Q_1\rangle$ depend non-monotonically on $\tau$. Both quantities follow a Lorentzian function, in agreement with Eq.\ref{avg}. We identify three regimes depending on $\tau$. In the adiabatic limit $\tau \gg \Gamma^{-1}$, the dynamics are reversible, the system remains in   equilibrium, and $\langle W_1\rangle \to0$. In the non-adiabatic regime $\tau \sim \Gamma^{-1}$, the dynamics are irreversible, the system in driven far from equilibrium, and $\langle W_1\rangle$ is maximized. In the limit $\tau \ll \Gamma^{-1}$, the dynamics are still non-equilibrium but  $\langle W_1\rangle \to0$. The work vanishes because the driving protocol $A(t)$ is so fast that $\alpha$ cannot respond to $A(t)$.

\section{Fluctuation Theorems} \label{S6}
\subsection{Symmetry functions} \label{S6.1}
	
While the second law demands $\langle W\rangle\geq0$, individual trajectories can yield $W<0$. At the heart of this possibility is the time-reversibility of microscopic dynamics. A solution to the OLE yielding $+W$ has a time-reversed counterpart yielding $-W$. However, the probabilities of observing $+W$ and $-W$ are not equal. The ratio of these probabilities is determined by a fluctuation theorem (FT).  Simply put, FTs are the extension of the second law to stochastic systems. They transform the inequality in the second law into an equality for the probability ratios of realizing positive and negative work, or positive and negative entropy production in general~\cite{Jarzynski11}.

FTs can generally be expressed in the form of a symmetry function~\cite{Ciliberto17},
\begin{equation} \label{symm}
	S\left(\frac{X_n}{\langle X_n\rangle}\right) =  \frac{k_BT}{\langle X_n\rangle}ln\left(\frac{\mathcal{P}\left(X_n/\langle X_n\rangle\right)}{\mathcal{P}\left(-X_n/\langle X_n\rangle\right)}\right).
\end{equation}
\noindent $X_n$ can represent  work $W_n$ or heat $Q_n$,  and $\langle X_n\rangle$ its average. $\mathcal{P}(X_n/\langle X_n\rangle)$ and $\mathcal{P}(-X_n/\langle X_n\rangle)$ are the probability of positive and negative $X_n/\langle X_n\rangle$, respectively. Thus, the symmetry function quantifies the asymmetry between the negative and positive regions of the PDF of $X_n$. Here, inspired by the works of Ciliberto~\cite{Douarche06} and Cohen~\cite{vanZon04_2} for mechanical and electrical oscillators, we calculate the symmetry functions of the work $W_n$ and heat $Q_n$ for our linear optical cavity driven on-resonance by a time-periodic laser amplitude.

\begin{figure*}[t]
	\includegraphics[]{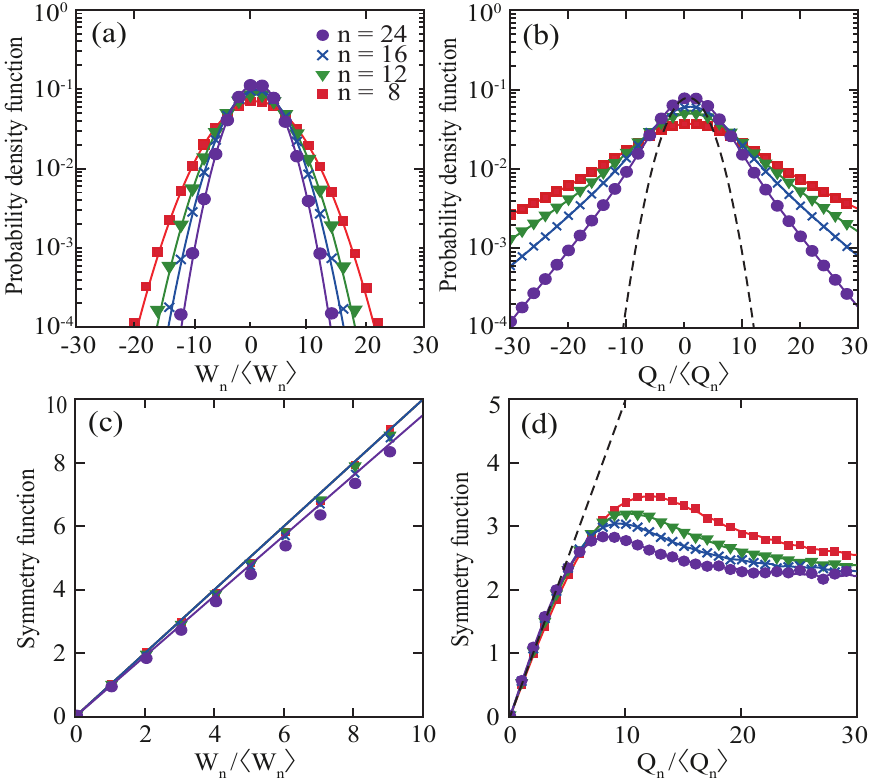}	
	\caption{Probability distribution functions of (a) the rescaled work $W_n/\langle W_n\rangle$ and (b) the rescaled heat $Q_n/\langle Q_n\rangle$, for varying number of cycles $n$ integrated over. Symbols represent numerical data. Solid curves are Gaussian fits in (a) and linear interpolation in (b). The dashed black curve in (b) is a Gaussian distribution fitted to the numerical data for $n=24$ in the neighbourhood of $Q_n/\langle Q_n\rangle=1$. Below their respective PDFs, the symmetry functions of $W_n/\langle W_n\rangle$ and $Q_n/\langle Q_n\rangle$ (from Eq.~\ref{symm}) are shown in (c) and (d). Model parameters are the same as  in Fig.~\ref{fig3}, except for $\tau= 100\Gamma^{-1}$ and $D=7\sqrt{\Gamma/2}$. }\label{fig4}
\end{figure*}

We focus on a particular class of FTs that describes non-equilibrium fluctuations around a steady state, the so-called Steady-State Fluctuating Theorem (SSFT)~\cite{Evans94,Evans02,Sevick08}. In terms of the symmetry functions, the SSFT predicts that $S(X_n/\langle X_n\rangle)=X_n/\langle X_n\rangle$ in the extensive limit $n\to\infty$. Essentially, the SSFT states that negative fluctuations of $X_n/\langle X_n\rangle$ are exponentially suppressed as $X_n/\langle X_n\rangle$ grows. At finite time the following linear relationship is assumed:
\begin{equation} \label{slope}
	S\left(\frac{X_n}{\langle X_n\rangle}\right) = \Sigma_X\frac{X_n}{\langle X_n\rangle}.
\end{equation}
\noindent The slope $\Sigma_X$ measures finite-time deviations from the SSFT. If $\Sigma_X=1$, the SSFT holds exactly~\cite{Joubaud07}.

We now elucidate the FT and symmetry functions through numerically-calculated PDFs of work and heat. These are presented in Figs.~\ref{fig4}(a) and ~\ref{fig4}(b), respectively. We obtained these PDFs from $10000$ trajectories of $\alpha_R$, each comprising $24$ cycles of a time-harmonic protocol in $A(t)$ with period $\tau=100\Gamma^{-1}$. We include 4 different PDFs in Figs.~\ref{fig4}(a,b),  corresponding to a different number $n$ of cycles over which the work or heat are calculated. PDFs are centred at $1$ because both heat and work are divided by  their average values.

Work distributions are Gaussian. Indeed, the solid curves Fig.~\ref{fig4}(a) are Gaussian distributions perfectly fitting the numerical data. Heat distributions, in contrast, are approximately Gaussian for small fluctuations only. To evidence this, in Fig.~\ref{fig4}(b) we fitted a Gaussian distribution to the numerical data for $n=24$ in the neighbourhood of $Q_n/\langle Q_n\rangle=1$. For $|Q_n/\langle Q_n\rangle| \gtrsim 5$, all heat PDFs are non-Gaussian.  Instead, they depend linearly on  $Q_n/\langle Q_n\rangle$ in the log-linear scale, meaning the distributions have exponential tails.

Notice that  both work and heat distributions become narrower as $n$, and therefore the integration time, increases. Accordingly, the probability of observing large negative fluctuations (the so-called ``transient violations of the second law'') decreases. Further, notice that large fluctuations are more likely in the heat than in the work. $\mathcal{P}(Q_n/\langle Q_n\rangle)$ is wider than $\mathcal{P}(W_n/\langle W_n\rangle)$ for all $n$.  This is due the fact that heat distributions have exponential tails, which fall off more slowly than the Gaussian distributions. The origin of these exponential tails is evident in Eq.~\ref{UWQ}. Unlike the work, the heat is nonlinear in $\alpha_R$. It contains higher order terms which make its PDF non-Gaussian.

Using Eq.~\ref{symm}, with $k_B T = \Gamma D^2/2$,
we can now calculate the symmetry functions of work and heat. These are presented in Figs.~\ref{fig4}(c) and ~\ref{fig4}(d), respectively.  The slope $\Sigma_W$ of the work symmetry function is equal to $1$ for any number of cycles integrated over. However, the heat symmetry function is very different. Firstly, the symmetry function of heat is only linear in the region of small $Q_n/\langle Q_n\rangle$. Therein, the slope $\Sigma_Q$ of the symmetry function is $0.5$. Secondly, for large fluctuations the symmetry function is nonlinear and converges to $\approx2$ for very large fluctuations. This means that large negative fluctuations in $Q_n$ are  still relevant compared to large positive fluctuations.
	
Overall, Figs.~\ref{fig4}(c) and ~\ref{fig4}(d) show that $W_n$ follows the standard SSFT exactly, while $Q_n$ does not. Instead, $Q_n$ follows an extended form of the SSFT developed by van Zon and Cohen~\cite{vanZon03}. The above statements hold when the driving conditions are adiabatic. Under non-adiabatic driving, deviations from the SSFT (and its extension) arise due to finite time effects, as shown next.

\subsection{Finite time corrections}  \label{S6.2}

    \begin{figure}[t]
		\includegraphics[]{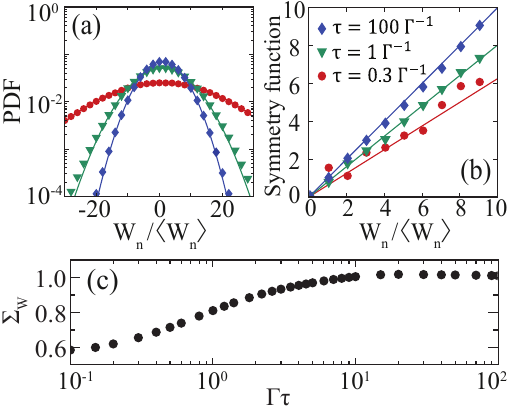}
		\caption{ (a) Probability distribution function of the rescaled work $W_n/\langle W_n\rangle$ under time-harmonic driving. The period $\tau$ is indicated in the legend of (b). Symbols are numerical data, solid curves are Gaussian fits. (b) Work symmetry functions using the same colour scheme. Solid lines are symmetry functions obtained from the Gaussian fits to the PDFs. (c) The slope $\Sigma_W$ of the symmetry function as a function of $\Gamma\tau$. Parameters are the same as in Fig.~\ref{fig4}, with $n = 8$.}\label{fig5}
	\end{figure}	

We now analyze finite time corrections to the SSFT when the period $\tau$ of the driving protocol becomes comparable to the equilibration time $\Gamma^{-1}$. We consider a fixed number of cycles $n=8$. In Fig.~\ref{fig5}(a) we compare PDFs of the rescaled work $W_n/\langle W_n\rangle$ for three distinct $\tau$, indicated in the legend of Fig.~\ref{fig5}(b). The PDFs are Gaussian for all $\tau$, as expected. They broaden as $\tau$ decreases. Large fluctuations become increasingly relevant in slow protocols.
	
Figure~\ref{fig5}(b) shows the work symmetry function for three different $\tau$. Notice how the slope $\Sigma_W$  of the symmetry function decreases as $\tau$ decreases. In Fig.~\ref{fig5}(c) we plot $\Sigma_W$ across a wide range of $\tau$.  $\Sigma_W$ is less than $1$ for small $\tau$, but converges to $1$ in the adiabatic limit $\tau \gg \Gamma^{-1}$.  The change in $\Sigma_W$ quantifies the finite time corrections to the SSFT for non-adiabatic driving conditions.

	\begin{figure}[t]
		\centering
		\includegraphics[]{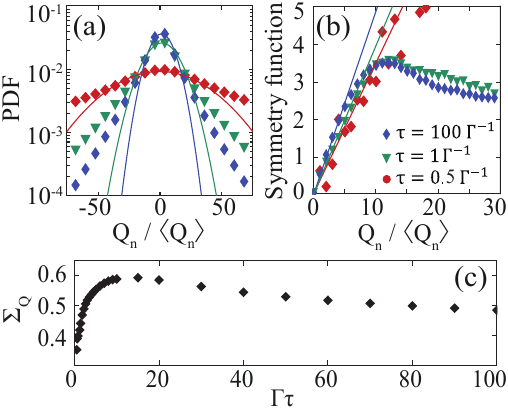}
		\caption{Same as in Fig.~\ref{fig5} but for the rescaled heat $Q_n/\langle Q_n\rangle$ instead of the work. Parameters are the same as in Fig.~\ref{fig5}.}\label{fig6}
	\end{figure}

Figure~\ref{fig6} presents a similar analysis to the one in Fig.~\ref{fig5}, but now for heat instead of work. Figure~\ref{fig6}(a) shows the PDFs. They are approximately Gaussian for small $\pm Q_n$, but exponential at the tails. As $\tau$ decreases the PDFs broaden and the Gaussian region widens. This significantly changes the symmetry function, as Fig.~\ref{fig6}(b) shows. In particular, the symmetry function becomes linear over a wider range of $Q_n/\langle Q_n\rangle$ for small $\tau$. In Fig.~\ref{fig6}(c) we plot the slope $\Sigma_Q$ of the heat symmetry function. We observe a non-monotonic dependence on $\tau$. $\Sigma_Q$ converges to $0.5$ in the adiabatic limit, consistent with the results in Fig.~\ref{fig4}.
    	
\subsection{Crook's fluctuation theorem} \label{S6.3}
	
We previously studied FTs for a protocol starting and ending in the same equilibrium steady state. Now we demonstrate a FT for the work done during the forward and backward parts of such a protocol; by forward and backward parts we mean the half-cycles whereby the laser amplitude increases and decreases, respectively. In particular, we demonstrate Crook's fluctuation theorem (CFT)~\cite{Crooks98} for our coherently-driven linear optical resonator. The CFT is a paradigm for understanding emergent phenomena~\cite{Collin05, England15, Ciliberto17}. It enables estimating free energy differences by measuring forward and backward work PDFs. Crucially, the CFT holds regardless of the speed of the process, and hence on how far from equilibrium the system is driven. However, the CFT assumes that the system starts and ends in equilibrium.

Consider a driving protocol that takes a system from initial to final state and back symmetrically. Then, the CFT states that
\begin{equation} \label{CFT}
	\frac{\mathcal{P}_f\left(W\right)}{\mathcal{P}_b\left(-W\right)} = e^{\beta \left(W-\delta F\right)}.
\end{equation}
\noindent $W$ is the work. $\mathcal{P}_f\left(W\right)$ is the probability of $+W$ being generated in the forward half-cycle and $\mathcal{P}_b\left(-W\right)$ is the probability of $-W$ being generated in the backward half-cycle. $\delta F$ is the free energy difference between initial and final states.

Essentially, Eq.~\ref{CFT} quantifies the reversibility of a transition between two equilibrium steady states. It does so in terms of the asymmetry of work distributions in the forward and backward directions. The crossing point of the two distributions, i.e. the value of $W$ for which $\mathcal{P}_f\left(W\right)$=$\mathcal{P}_b\left(-W\right)$,  is the exactly $\delta F$. This possibility, namely to estimate equilibrium free energies  by performing non-equilibrium measurements, is possibly the main reason for which the CFT became a pillar of stochastic thermodynamics.
	
	\begin{figure}[t]
		\includegraphics[]{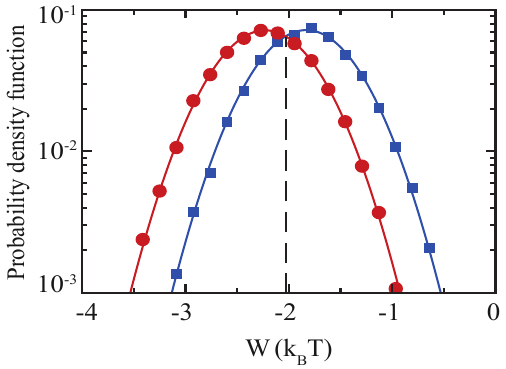}
		\caption{ Blue squares and red dots indicate probability distributions of the work done during the forward and backward half-cycles along a time-harmonic protocol in the laser amplitude, respectively. Solid curves are Gaussian fits. The vertical dashed line indicates the intersection point of the PDFs, which marks the free energy difference between the initial and final states according to Eq.~\ref{CFT}. Model parameters are same as in Fig.~\ref{fig3} except: $\tau = 200\Gamma^{-1}$, $A_0=5\sqrt{\Gamma}$, $D=7\sqrt{\Gamma/2}$.}\label{fig7}
	\end{figure}
	
The CFT has been used to measure free energy differences in single molecules~\cite{Alemany15,Collin05,Liphardt02,Hummer01} and mechanical systems~\cite{Douarche05}. Here we use it in the context of our coherently-driven linear optical resonator.  To this end, we performed numerical simulations of Eq.~\ref{OLE1D} with a time-harmonic protocol in the laser amplitude. $A$ increases from 0 to $10\sqrt{\Gamma}$ in the forward part of the protocol, and decreases from $10\sqrt{\Gamma}$ to $0$ in the backward part. The period is $\tau=200\Gamma^{-1}$. For each half-cycle we calculated the work done using Eq.~\ref{UWQb}. Finally, we obtained the distributions $\mathcal{P}_f\left(W\right)$ and $\mathcal{P}_b\left(-W\right)$ from an ensemble of $10000$ independent cycles.

Figure~\ref{fig7} shows the forward and backward work distribution as red and blue dots, respectively. The two distributions intersect at $W=-2.041$ $k_BT$, as indicated by the dashed line in Fig.~\ref{fig7}. According to the CFT, this is exactly the free energy difference between the steady states at the start and end of our protocol in $A$. We can verify this result by calculating free energies of those states using Eq.~\ref{F}. Indeed, inserting the parameters reported in Fig.~\ref{fig7} into Eq.~\ref{F}, we exactly calculate the free energy difference between initial and final states of our protocol to be $\delta F=-2.041$ $k_BT$. We highlight that while the results presented in this manuscript were obtained for a large driving period, we verified that the results are independent of the period. However, if a small (compared to $\Gamma^{-1}$) driving period is used, the system needs to be allowed to equilibrate by halting the protocol for some time at the start and end of each half-cycle.

\section{Conclusion and perspectives} \label{S7}

In summary, we presented a complete framework of stochastic thermodynamics for a single-mode linear optical cavity driven on resonance. We showed that light in such a cavity displays effective equilibrium behaviour. We formulated the first and second laws of thermodynamics for the cavity in terms of optical control parameters and observables. Next, we analysed the work and heat generated when the cavity is driven by a periodically-modulated laser amplitude. The averaged work and heat produced per cycle is maximized when the driving period is commensurate with the equilibration time of the system and the dynamics are strongly irreversible. Further,  we discussed fluctuation theorems for work and heat, including their finite time corrections. Finally, we showed how measurements of forward and backward work can enable the estimation of free energy differences between optical states via Crook's fluctuation theorem.

Our work opens a new research avenue at the crossroad between stochastic thermodynamics and nanophotonics. To date, nanophotonics has primarily provided tools for probing stochastic thermodynamics of material systems in new regimes. A prime example of this is in the burgeoning field of levitodynamics, where sophisticated nanophotonic methods are used to trap particles in new settings~\cite{Melo24}, or to introduce entirely new types of particles in the trap~\cite{Lepeshov23}.  Here, in contrast, we completely reversed the role of light and matter: the trap is made of matter, while light is the stochastic thermodynamic system.  This difference opens intriguing opportunities for new fundamental physics studies and technological applications.

Fundamentally, optical cavities can facilitate probing fluctuation theorems for systems transitioning from one non-equilibrium steady state (NESS) to another~\cite{Chernyak06, Seifert12}. That is challenging to accomplish in material systems, which tend to relax to equilibrium steady states. In a laser-driven cavity, in contrast, one simply needs to shift the laser-cavity detuning away from zero for the system to settle in a NESS. For non-zero detuning, the phase space dynamics of light in a single-mode cavity is formally equivalent to the two-dimensional non-equilibrium motion of a Brownian particle in a stirred fluid~\cite{Ramesh24}. A second fundamentally-interesting extension of our results could be to insert a thermo-optical nonlinear medium in the cavity~\cite{Geng20, Peters21}. This can enable probing fluctuation theorems in non-Markovian regimes, where analytical results are hard to obtain and experiments are particularly valuable.  A third and fundamentally-interesting feature of optical cavities in the context of stochastic thermodynamics is the extremely wide dynamic range that can be easily probed. For example, a single-mode cavity with Kerr nonlinearity can take longer than the age of the universe to relax to its steady state~\cite{Rodriguez17, Casteels17}. Meanwhile, relevant dynamics unfold within the dissipation time, typically on the order of a
picosecond. Thus, within a single second one can in principle (pending practical limitations) attain statistics of dynamics spanning 12 orders of magnitude in time. No other system can give access to such a wide dynamic range and with such ease. That is valuable for studying rare events, which are the most interesting and challenging to detect when probing stochastic phenomena.

Finally, we foresee exciting technological  opportunities in the extension of stochastic thermodynamic concepts to optical cavities. For instance, optimal protocols could be designed to drive an optical system from one state to another with minimum dissipation~\cite{Tafoya19}. Alternatively, time-information uncertainty relations~\cite{Nicholson20} could be used to establish speed limits for transitions between optical states, and thus for optimizing optical devices using such transitions. Many fascinating opportunities emerge from the recognition that resonant optical systems can be described, and eventually optimized, like light engines.

\begin{acknowledgement}
\noindent This work is part of the research programme of the Netherlands Organisation for Scientific Research (NWO). We thank Nicola Carlon Zambon, Sergio Ciliberto, and Christopher Jarzynski for stimulating discussions. S.R.K.R. acknowledges an ERC Starting Grant with project number 852694.
\end{acknowledgement}
	

\begin{mcitethebibliography}{56}
\providecommand*\natexlab[1]{#1}
\providecommand*\mciteSetBstSublistMode[1]{}
\providecommand*\mciteSetBstMaxWidthForm[2]{}
\providecommand*\mciteBstWouldAddEndPuncttrue
  {\def\EndOfBibitem{\unskip.}}
\providecommand*\mciteBstWouldAddEndPunctfalse
  {\let\EndOfBibitem\relax}
\providecommand*\mciteSetBstMidEndSepPunct[3]{}
\providecommand*\mciteSetBstSublistLabelBeginEnd[3]{}
\providecommand*\EndOfBibitem{}
\mciteSetBstSublistMode{f}
\mciteSetBstMaxWidthForm{subitem}{(\alph{mcitesubitemcount})}
\mciteSetBstSublistLabelBeginEnd
  {\mcitemaxwidthsubitemform\space}
  {\relax}
  {\relax}

\bibitem[Jarzynski(2011)]{Jarzynski11}
Jarzynski,~C. Equalities and Inequalities: Irreversibility and the Second Law
  of Thermodynamics at the Nanoscale. \emph{Annu. Rev. Condens. Matter Phys.}
  \textbf{2011}, \emph{2}, 329--351\relax
\mciteBstWouldAddEndPuncttrue
\mciteSetBstMidEndSepPunct{\mcitedefaultmidpunct}
{\mcitedefaultendpunct}{\mcitedefaultseppunct}\relax
\EndOfBibitem
\bibitem[Seifert(2012)]{Seifert12}
Seifert,~U. Stochastic thermodynamics, fluctuation theorems and molecular
  machines. \emph{Rep. Prog. Phys.} \textbf{2012}, \emph{75}, 126001\relax
\mciteBstWouldAddEndPuncttrue
\mciteSetBstMidEndSepPunct{\mcitedefaultmidpunct}
{\mcitedefaultendpunct}{\mcitedefaultseppunct}\relax
\EndOfBibitem
\bibitem[Parrondo \latin{et~al.}(2015)Parrondo, Horowitz, and
  Sagawa]{Parrondo15}
Parrondo,~J.~M.; Horowitz,~J.~M.; Sagawa,~T. Thermodynamics of information.
  \emph{Nat. Phys.} \textbf{2015}, \emph{11}, 131--139\relax
\mciteBstWouldAddEndPuncttrue
\mciteSetBstMidEndSepPunct{\mcitedefaultmidpunct}
{\mcitedefaultendpunct}{\mcitedefaultseppunct}\relax
\EndOfBibitem
\bibitem[Ciliberto(2017)]{Ciliberto17}
Ciliberto,~S. Experiments in Stochastic Thermodynamics: Short History and
  Perspectives. \emph{Phys. Rev. X} \textbf{2017}, \emph{7}, 021051\relax
\mciteBstWouldAddEndPuncttrue
\mciteSetBstMidEndSepPunct{\mcitedefaultmidpunct}
{\mcitedefaultendpunct}{\mcitedefaultseppunct}\relax
\EndOfBibitem
\bibitem[Blickle and Bechinger(2012)Blickle, and Bechinger]{Blickle12}
Blickle,~V.; Bechinger,~C. Realization of a micrometre-sized stochastic heat
  engine. \emph{Nat. Phys.} \textbf{2012}, \emph{8}, 143--146\relax
\mciteBstWouldAddEndPuncttrue
\mciteSetBstMidEndSepPunct{\mcitedefaultmidpunct}
{\mcitedefaultendpunct}{\mcitedefaultseppunct}\relax
\EndOfBibitem
\bibitem[Mart{\'\i}nez \latin{et~al.}(2016)Mart{\'\i}nez, Rold{\'a}n, Dinis,
  Petrov, Parrondo, and Rica]{Martinez16}
Mart{\'\i}nez,~I.~A.; Rold{\'a}n,~{\'E}.; Dinis,~L.; Petrov,~D.;
  Parrondo,~J.~M.; Rica,~R.~A. Brownian carnot engine. \emph{Nat. Phys.}
  \textbf{2016}, \emph{12}, 67--70\relax
\mciteBstWouldAddEndPuncttrue
\mciteSetBstMidEndSepPunct{\mcitedefaultmidpunct}
{\mcitedefaultendpunct}{\mcitedefaultseppunct}\relax
\EndOfBibitem
\bibitem[Klaers \latin{et~al.}(2010)Klaers, Vewinger, and Weitz]{Klaers10T}
Klaers,~J.; Vewinger,~F.; Weitz,~M. Thermalization of a two-dimensional
  photonic gas in a ‘white wall’ photon box. \emph{Nat. Phys.}
  \textbf{2010}, \emph{6}, 512--515\relax
\mciteBstWouldAddEndPuncttrue
\mciteSetBstMidEndSepPunct{\mcitedefaultmidpunct}
{\mcitedefaultendpunct}{\mcitedefaultseppunct}\relax
\EndOfBibitem
\bibitem[Thomas \latin{et~al.}(2018)Thomas, Dhakal, Raza, Peyskens, and
  Baets]{LeThomas18}
Thomas,~N.~L.; Dhakal,~A.; Raza,~A.; Peyskens,~F.; Baets,~R. Impact of
  fundamental thermodynamic fluctuations on light propagating in photonic
  waveguides made of amorphous materials. \emph{Optica} \textbf{2018},
  \emph{5}, 328--336\relax
\mciteBstWouldAddEndPuncttrue
\mciteSetBstMidEndSepPunct{\mcitedefaultmidpunct}
{\mcitedefaultendpunct}{\mcitedefaultseppunct}\relax
\EndOfBibitem
\bibitem[Miller and Anders(2018)Miller, and Anders]{Miller18}
Miller,~H.~J.; Anders,~J. Energy-temperature uncertainty relation in quantum
  thermodynamics. \emph{Nat. Commun.} \textbf{2018}, \emph{9}, 2203\relax
\mciteBstWouldAddEndPuncttrue
\mciteSetBstMidEndSepPunct{\mcitedefaultmidpunct}
{\mcitedefaultendpunct}{\mcitedefaultseppunct}\relax
\EndOfBibitem
\bibitem[Mann \latin{et~al.}(2019)Mann, Sounas, and Al\`{u}]{Mann19}
Mann,~S.~A.; Sounas,~D.~L.; Al\`{u},~A. Nonreciprocal cavities and the
  time--bandwidth limit. \emph{Optica} \textbf{2019}, \emph{6}, 104--110\relax
\mciteBstWouldAddEndPuncttrue
\mciteSetBstMidEndSepPunct{\mcitedefaultmidpunct}
{\mcitedefaultendpunct}{\mcitedefaultseppunct}\relax
\EndOfBibitem
\bibitem[Panuski \latin{et~al.}(2020)Panuski, Englund, and Hamerly]{Hamerly20}
Panuski,~C.; Englund,~D.; Hamerly,~R. Fundamental Thermal Noise Limits for
  Optical Microcavities. \emph{Phys. Rev. X} \textbf{2020}, \emph{10},
  041046\relax
\mciteBstWouldAddEndPuncttrue
\mciteSetBstMidEndSepPunct{\mcitedefaultmidpunct}
{\mcitedefaultendpunct}{\mcitedefaultseppunct}\relax
\EndOfBibitem
\bibitem[Sanders \latin{et~al.}(2021)Sanders, Zundel, Kort-Kamp, Dalvit, and
  Manjavacas]{Manjavacas21}
Sanders,~S.; Zundel,~L.; Kort-Kamp,~W. J.~M.; Dalvit,~D. A.~R.; Manjavacas,~A.
  Near-Field Radiative Heat Transfer Eigenmodes. \emph{Phys. Rev. Lett.}
  \textbf{2021}, \emph{126}, 193601\relax
\mciteBstWouldAddEndPuncttrue
\mciteSetBstMidEndSepPunct{\mcitedefaultmidpunct}
{\mcitedefaultendpunct}{\mcitedefaultseppunct}\relax
\EndOfBibitem
\bibitem[Fan and Li(2022)Fan, and Li]{Fan22}
Fan,~S.; Li,~W. Photonics and thermodynamics concepts in radiative cooling.
  \emph{Nat. Photonics} \textbf{2022}, \emph{16}, 182--190\relax
\mciteBstWouldAddEndPuncttrue
\mciteSetBstMidEndSepPunct{\mcitedefaultmidpunct}
{\mcitedefaultendpunct}{\mcitedefaultseppunct}\relax
\EndOfBibitem
\bibitem[Hassani~Gangaraj and Monticone(2022)Hassani~Gangaraj, and
  Monticone]{Monticone22}
Hassani~Gangaraj,~S.~A.; Monticone,~F. Drifting Electrons: Nonreciprocal
  Plasmonics and Thermal Photonics. \emph{ACS Photonics} \textbf{2022},
  \emph{9}, 806--819\relax
\mciteBstWouldAddEndPuncttrue
\mciteSetBstMidEndSepPunct{\mcitedefaultmidpunct}
{\mcitedefaultendpunct}{\mcitedefaultseppunct}\relax
\EndOfBibitem
\bibitem[Wu \latin{et~al.}(2019)Wu, Hassan, and Christodoulides]{Wu2019}
Wu,~F.~O.; Hassan,~A.~U.; Christodoulides,~D.~N. Thermodynamic theory of highly
  multimoded nonlinear optical systems. \emph{Nat. Photonics} \textbf{2019},
  \emph{13}, 776--782\relax
\mciteBstWouldAddEndPuncttrue
\mciteSetBstMidEndSepPunct{\mcitedefaultmidpunct}
{\mcitedefaultendpunct}{\mcitedefaultseppunct}\relax
\EndOfBibitem
\bibitem[Muniz \latin{et~al.}(2023)Muniz, Wu, Jung, Khajavikhan,
  Christodoulides, and Peschel]{Muniz23}
Muniz,~A. L.~M.; Wu,~F.~O.; Jung,~P.~S.; Khajavikhan,~M.;
  Christodoulides,~D.~N.; Peschel,~U. Observation of photon-photon
  thermodynamic processes under negative optical temperature conditions.
  \emph{Science} \textbf{2023}, \emph{379}, 1019--1023\relax
\mciteBstWouldAddEndPuncttrue
\mciteSetBstMidEndSepPunct{\mcitedefaultmidpunct}
{\mcitedefaultendpunct}{\mcitedefaultseppunct}\relax
\EndOfBibitem
\bibitem[Pyrialakos \latin{et~al.}(2022)Pyrialakos, Ren, Jung, Khajavikhan, and
  Christodoulides]{Pyrialakos22}
Pyrialakos,~G.~G.; Ren,~H.; Jung,~P.~S.; Khajavikhan,~M.;
  Christodoulides,~D.~N. Thermalization Dynamics of Nonlinear Non-Hermitian
  Optical Lattices. \emph{Phys. Rev. Lett.} \textbf{2022}, \emph{128},
  213901\relax
\mciteBstWouldAddEndPuncttrue
\mciteSetBstMidEndSepPunct{\mcitedefaultmidpunct}
{\mcitedefaultendpunct}{\mcitedefaultseppunct}\relax
\EndOfBibitem
\bibitem[Kewming and Shrapnel(2022)Kewming, and Shrapnel]{Kewming22}
Kewming,~M.~J.; Shrapnel,~S. Entropy production and fluctuation theorems in a
  continuously monitored optical cavity at zero temperature. \emph{{Quantum}}
  \textbf{2022}, \emph{6}, 685\relax
\mciteBstWouldAddEndPuncttrue
\mciteSetBstMidEndSepPunct{\mcitedefaultmidpunct}
{\mcitedefaultendpunct}{\mcitedefaultseppunct}\relax
\EndOfBibitem
\bibitem[Mader \latin{et~al.}(2022)Mader, Benedikter, Husel, Hänsch, and
  Hunger]{Mader22}
Mader,~M.; Benedikter,~J.; Husel,~L.; Hänsch,~T.~W.; Hunger,~D. Quantitative
  Determination of the Complex Polarizability of Individual Nanoparticles by
  Scanning Cavity Microscopy. \emph{ACS Photonics} \textbf{2022}, \emph{9},
  466--473\relax
\mciteBstWouldAddEndPuncttrue
\mciteSetBstMidEndSepPunct{\mcitedefaultmidpunct}
{\mcitedefaultendpunct}{\mcitedefaultseppunct}\relax
\EndOfBibitem
\bibitem[Houghton \latin{et~al.}(2024)Houghton, Kashanian, Derrien, Masuda, and
  Vollmer]{Vollmer24}
Houghton,~M.~C.; Kashanian,~S.~V.; Derrien,~T.~L.; Masuda,~K.; Vollmer,~F.
  Whispering-Gallery Mode Optoplasmonic Microcavities: From Advanced
  Single-Molecule Sensors and Microlasers to Applications in Synthetic Biology.
  \emph{ACS Photonics} \textbf{2024}, \emph{11}, 892--903\relax
\mciteBstWouldAddEndPuncttrue
\mciteSetBstMidEndSepPunct{\mcitedefaultmidpunct}
{\mcitedefaultendpunct}{\mcitedefaultseppunct}\relax
\EndOfBibitem
\bibitem[Perrier \latin{et~al.}(2020)Perrier, Greveling, Wouters, Rodriguez,
  Lehoucq, Combri\'{e}, de~Rossi, Faez, and Mosk]{Perrier20}
Perrier,~K.; Greveling,~S.; Wouters,~H.; Rodriguez,~S. R.~K.; Lehoucq,~G.;
  Combri\'{e},~S.; de~Rossi,~A.; Faez,~S.; Mosk,~A.~P. Thermo-optical dynamics
  of a nonlinear GaInP photonic crystal nanocavity depend on the optical mode
  profile. \emph{OSA Contin.} \textbf{2020}, \emph{3}, 1879--1890\relax
\mciteBstWouldAddEndPuncttrue
\mciteSetBstMidEndSepPunct{\mcitedefaultmidpunct}
{\mcitedefaultendpunct}{\mcitedefaultseppunct}\relax
\EndOfBibitem
\bibitem[Li \latin{et~al.}(2024)Li, Li, Moon, Wilmington, Lin, Gundogdu, and
  Gu]{Li24}
Li,~X.; Li,~J.; Moon,~J.; Wilmington,~R.~L.; Lin,~D.; Gundogdu,~K.; Gu,~Q.
  Experimental Observation of Purcell-Enhanced Spontaneous Emission in a
  Single-Mode Plasmonic Nanocavity. \emph{ACS Photonics} \textbf{2024},
  \emph{XXXX}, XXX--XXX\relax
\mciteBstWouldAddEndPuncttrue
\mciteSetBstMidEndSepPunct{\mcitedefaultmidpunct}
{\mcitedefaultendpunct}{\mcitedefaultseppunct}\relax
\EndOfBibitem
\bibitem[Ramesh \latin{et~al.}(2024)Ramesh, Peters, and Rodriguez]{Ramesh24}
Ramesh,~V.~G.; Peters,~K. J.~H.; Rodriguez,~S. R.~K. Arcsine Laws of Light.
  \emph{Phys. Rev. Lett.} \textbf{2024}, \emph{132}, 133801\relax
\mciteBstWouldAddEndPuncttrue
\mciteSetBstMidEndSepPunct{\mcitedefaultmidpunct}
{\mcitedefaultendpunct}{\mcitedefaultseppunct}\relax
\EndOfBibitem
\bibitem[Peters \latin{et~al.}(2023)Peters, Busink, Ackermans, Cogn\'ee, and
  Rodriguez]{Peters23}
Peters,~K. J.~H.; Busink,~J.; Ackermans,~P.; Cogn\'ee,~K.~G.; Rodriguez,~S.
  R.~K. Scalar potentials for light in a cavity. \emph{Phys. Rev. Res.}
  \textbf{2023}, \emph{5}, 013154\relax
\mciteBstWouldAddEndPuncttrue
\mciteSetBstMidEndSepPunct{\mcitedefaultmidpunct}
{\mcitedefaultendpunct}{\mcitedefaultseppunct}\relax
\EndOfBibitem
\bibitem[Chernyak \latin{et~al.}(2006)Chernyak, Chertkov, and
  Jarzynski]{Chernyak06}
Chernyak,~V.~Y.; Chertkov,~M.; Jarzynski,~C. Path-integral analysis of
  fluctuation theorems for general Langevin processes. \emph{J. Stat. Mech.}
  \textbf{2006}, \emph{2006}, P08001\relax
\mciteBstWouldAddEndPuncttrue
\mciteSetBstMidEndSepPunct{\mcitedefaultmidpunct}
{\mcitedefaultendpunct}{\mcitedefaultseppunct}\relax
\EndOfBibitem
\bibitem[Kiesewetter \latin{et~al.}(2016)Kiesewetter, Polkinghorne, Opanchuk,
  and Drummond]{xspde}
Kiesewetter,~S.; Polkinghorne,~R.; Opanchuk,~B.; Drummond,~P.~D. {xSPDE}:
  Extensible software for stochastic equations. \emph{{SoftwareX}}
  \textbf{2016}, \emph{5}, 12--15\relax
\mciteBstWouldAddEndPuncttrue
\mciteSetBstMidEndSepPunct{\mcitedefaultmidpunct}
{\mcitedefaultendpunct}{\mcitedefaultseppunct}\relax
\EndOfBibitem
\bibitem[Sekimoto(1998)]{Sekimoto98}
Sekimoto,~K. {Langevin Equation and Thermodynamics}. \emph{Prog. Theor. Phys.}
  \textbf{1998}, \emph{130}, 17--27\relax
\mciteBstWouldAddEndPuncttrue
\mciteSetBstMidEndSepPunct{\mcitedefaultmidpunct}
{\mcitedefaultendpunct}{\mcitedefaultseppunct}\relax
\EndOfBibitem
\bibitem[Jarzynski(1997)]{Jarzynski97}
Jarzynski,~C. Nonequilibrium Equality for Free Energy Differences. \emph{Phys.
  Rev. Lett.} \textbf{1997}, \emph{78}, 2690--2693\relax
\mciteBstWouldAddEndPuncttrue
\mciteSetBstMidEndSepPunct{\mcitedefaultmidpunct}
{\mcitedefaultendpunct}{\mcitedefaultseppunct}\relax
\EndOfBibitem
\bibitem[Hummer and Szabo(2001)Hummer, and Szabo]{Hummer01}
Hummer,~G.; Szabo,~A. Free energy reconstruction from nonequilibrium
  single-molecule pulling experiments. \emph{Proc. Natl. Acad. Sci. U.S.A.}
  \textbf{2001}, \emph{98}, 3658--3661\relax
\mciteBstWouldAddEndPuncttrue
\mciteSetBstMidEndSepPunct{\mcitedefaultmidpunct}
{\mcitedefaultendpunct}{\mcitedefaultseppunct}\relax
\EndOfBibitem
\bibitem[Douarche \latin{et~al.}(2005)Douarche, Ciliberto, and
  Petrosyan]{Douarche05}
Douarche,~F.; Ciliberto,~S.; Petrosyan,~A. Estimate of the free energy
  difference in mechanical systems from work fluctuations: experiments and
  models. \emph{J. Stat. Mech.: Theory Exp.} \textbf{2005}, \emph{2005},
  P09011\relax
\mciteBstWouldAddEndPuncttrue
\mciteSetBstMidEndSepPunct{\mcitedefaultmidpunct}
{\mcitedefaultendpunct}{\mcitedefaultseppunct}\relax
\EndOfBibitem
\bibitem[Wang \latin{et~al.}(2002)Wang, Sevick, Mittag, Searles, and
  Evans]{Wang02}
Wang,~G.~M.; Sevick,~E.~M.; Mittag,~E.; Searles,~D.~J.; Evans,~D.~J.
  Experimental Demonstration of Violations of the Second Law of Thermodynamics
  for Small Systems and Short Time Scales. \emph{Phys. Rev. Lett.}
  \textbf{2002}, \emph{89}, 050601\relax
\mciteBstWouldAddEndPuncttrue
\mciteSetBstMidEndSepPunct{\mcitedefaultmidpunct}
{\mcitedefaultendpunct}{\mcitedefaultseppunct}\relax
\EndOfBibitem
\bibitem[Ritort(2004)]{Ritort04}
Ritort,~F. Work fluctuations, transient violations of the second law and
  free-energy recovery methods: Perspectives in theory and experiments.
  \emph{Progress in Mathematical Physics} \textbf{2004}, \emph{38}, 193\relax
\mciteBstWouldAddEndPuncttrue
\mciteSetBstMidEndSepPunct{\mcitedefaultmidpunct}
{\mcitedefaultendpunct}{\mcitedefaultseppunct}\relax
\EndOfBibitem
\bibitem[Wang \latin{et~al.}(2005)Wang, Reid, Carberry, Williams, Sevick, and
  Evans]{Wang05}
Wang,~G.~M.; Reid,~J.~C.; Carberry,~D.~M.; Williams,~D. R.~M.; Sevick,~E.~M.;
  Evans,~D.~J. Experimental study of the fluctuation theorem in a
  nonequilibrium steady state. \emph{Phys. Rev. E} \textbf{2005}, \emph{71},
  046142\relax
\mciteBstWouldAddEndPuncttrue
\mciteSetBstMidEndSepPunct{\mcitedefaultmidpunct}
{\mcitedefaultendpunct}{\mcitedefaultseppunct}\relax
\EndOfBibitem
\bibitem[Blickle \latin{et~al.}(2006)Blickle, Speck, Helden, Seifert, and
  Bechinger]{Blickle06}
Blickle,~V.; Speck,~T.; Helden,~L.; Seifert,~U.; Bechinger,~C. Thermodynamics
  of a Colloidal Particle in a Time-Dependent Nonharmonic Potential.
  \emph{Phys. Rev. Lett.} \textbf{2006}, \emph{96}, 070603\relax
\mciteBstWouldAddEndPuncttrue
\mciteSetBstMidEndSepPunct{\mcitedefaultmidpunct}
{\mcitedefaultendpunct}{\mcitedefaultseppunct}\relax
\EndOfBibitem
\bibitem[Schuler \latin{et~al.}(2005)Schuler, Speck, Tietz, Wrachtrup, and
  Seifert]{Schuler05}
Schuler,~S.; Speck,~T.; Tietz,~C.; Wrachtrup,~J.; Seifert,~U. Experimental Test
  of the Fluctuation Theorem for a Driven Two-Level System with Time-Dependent
  Rates. \emph{Phys. Rev. Lett.} \textbf{2005}, \emph{94}, 180602\relax
\mciteBstWouldAddEndPuncttrue
\mciteSetBstMidEndSepPunct{\mcitedefaultmidpunct}
{\mcitedefaultendpunct}{\mcitedefaultseppunct}\relax
\EndOfBibitem
\bibitem[Joubaud \latin{et~al.}(2007)Joubaud, Garnier, and
  Ciliberto]{Joubaud07}
Joubaud,~S.; Garnier,~N.~B.; Ciliberto,~S. Fluctuation theorems for harmonic
  oscillators. \emph{J. Stat. Mech} \textbf{2007}, \emph{2007}, P09018\relax
\mciteBstWouldAddEndPuncttrue
\mciteSetBstMidEndSepPunct{\mcitedefaultmidpunct}
{\mcitedefaultendpunct}{\mcitedefaultseppunct}\relax
\EndOfBibitem
\bibitem[Douarche \latin{et~al.}(2006)Douarche, Joubaud, Garnier, Petrosyan,
  and Ciliberto]{Douarche06}
Douarche,~F.; Joubaud,~S.; Garnier,~N.~B.; Petrosyan,~A.; Ciliberto,~S. Work
  Fluctuation Theorems for Harmonic Oscillators. \emph{Phys. Rev. Lett.}
  \textbf{2006}, \emph{97}, 140603\relax
\mciteBstWouldAddEndPuncttrue
\mciteSetBstMidEndSepPunct{\mcitedefaultmidpunct}
{\mcitedefaultendpunct}{\mcitedefaultseppunct}\relax
\EndOfBibitem
\bibitem[van Zon \latin{et~al.}(2004)van Zon, Ciliberto, and Cohen]{vanZon04_2}
van Zon,~R.; Ciliberto,~S.; Cohen,~E. G.~D. Power and Heat Fluctuation Theorems
  for Electric Circuits. \emph{Phys. Rev. Lett.} \textbf{2004}, \emph{92},
  130601\relax
\mciteBstWouldAddEndPuncttrue
\mciteSetBstMidEndSepPunct{\mcitedefaultmidpunct}
{\mcitedefaultendpunct}{\mcitedefaultseppunct}\relax
\EndOfBibitem
\bibitem[Evans and Searles(1994)Evans, and Searles]{Evans94}
Evans,~D.~J.; Searles,~D.~J. Equilibrium microstates which generate second law
  violating steady states. \emph{Phys. Rev. E} \textbf{1994}, \emph{50},
  1645--1648\relax
\mciteBstWouldAddEndPuncttrue
\mciteSetBstMidEndSepPunct{\mcitedefaultmidpunct}
{\mcitedefaultendpunct}{\mcitedefaultseppunct}\relax
\EndOfBibitem
\bibitem[Evans and Searles(2002)Evans, and Searles]{Evans02}
Evans,~D.~J.; Searles,~D.~J. The Fluctuation Theorem. \emph{Adv. Phys.}
  \textbf{2002}, \emph{51}, 1529--1585\relax
\mciteBstWouldAddEndPuncttrue
\mciteSetBstMidEndSepPunct{\mcitedefaultmidpunct}
{\mcitedefaultendpunct}{\mcitedefaultseppunct}\relax
\EndOfBibitem
\bibitem[Sevick \latin{et~al.}(2008)Sevick, Prabhakar, Williams, and
  Searles]{Sevick08}
Sevick,~E.; Prabhakar,~R.; Williams,~S.~R.; Searles,~D.~J. Fluctuation
  Theorems. \emph{Annu. Rev. Phys. Chem.} \textbf{2008}, \emph{59},
  603--633\relax
\mciteBstWouldAddEndPuncttrue
\mciteSetBstMidEndSepPunct{\mcitedefaultmidpunct}
{\mcitedefaultendpunct}{\mcitedefaultseppunct}\relax
\EndOfBibitem
\bibitem[van Zon and Cohen(2003)van Zon, and Cohen]{vanZon03}
van Zon,~R.; Cohen,~E. G.~D. Extension of the Fluctuation Theorem. \emph{Phys.
  Rev. Lett.} \textbf{2003}, \emph{91}, 110601\relax
\mciteBstWouldAddEndPuncttrue
\mciteSetBstMidEndSepPunct{\mcitedefaultmidpunct}
{\mcitedefaultendpunct}{\mcitedefaultseppunct}\relax
\EndOfBibitem
\bibitem[Crooks(1998)]{Crooks98}
Crooks,~G.~E. Nonequilibrium Measurements of Free Energy Differences for
  Microscopically Reversible Markovian Systems. \emph{J. Stat. Phys.}
  \textbf{1998}, \emph{90}, 1481--1487\relax
\mciteBstWouldAddEndPuncttrue
\mciteSetBstMidEndSepPunct{\mcitedefaultmidpunct}
{\mcitedefaultendpunct}{\mcitedefaultseppunct}\relax
\EndOfBibitem
\bibitem[Collin \latin{et~al.}(2005)Collin, Ritort, Jarzynski, Smith, Tinoco,
  and Bustamante]{Collin05}
Collin,~D.; Ritort,~F.; Jarzynski,~C.; Smith,~S.~B.; Tinoco,~I.; Bustamante,~C.
  Verification of the Crooks fluctuation theorem and recovery of RNA folding
  free energies. \emph{Nature} \textbf{2005}, \emph{437}, 231--234\relax
\mciteBstWouldAddEndPuncttrue
\mciteSetBstMidEndSepPunct{\mcitedefaultmidpunct}
{\mcitedefaultendpunct}{\mcitedefaultseppunct}\relax
\EndOfBibitem
\bibitem[England(2015)]{England15}
England,~J.~L. Dissipative adaptation in driven self-assembly. \emph{Nat.
  Nanotechnol.} \textbf{2015}, \emph{10}, 919--923\relax
\mciteBstWouldAddEndPuncttrue
\mciteSetBstMidEndSepPunct{\mcitedefaultmidpunct}
{\mcitedefaultendpunct}{\mcitedefaultseppunct}\relax
\EndOfBibitem
\bibitem[Alemany \latin{et~al.}(2015)Alemany, Ribezzi-Crivellari, and
  Ritort]{Alemany15}
Alemany,~A.; Ribezzi-Crivellari,~M.; Ritort,~F. From free energy measurements
  to thermodynamic inference in nonequilibrium small systems. \emph{New J.
  Phys.} \textbf{2015}, \emph{17}, 075009\relax
\mciteBstWouldAddEndPuncttrue
\mciteSetBstMidEndSepPunct{\mcitedefaultmidpunct}
{\mcitedefaultendpunct}{\mcitedefaultseppunct}\relax
\EndOfBibitem
\bibitem[Liphardt \latin{et~al.}(2002)Liphardt, Dumont, Smith, Tinoco, and
  Bustamante]{Liphardt02}
Liphardt,~J.; Dumont,~S.; Smith,~S.~B.; Tinoco,~I.; Bustamante,~C. Equilibrium
  Information from Nonequilibrium Measurements in an Experimental Test of
  Jarzynski's Equality. \emph{Science} \textbf{2002}, \emph{296},
  1832--1835\relax
\mciteBstWouldAddEndPuncttrue
\mciteSetBstMidEndSepPunct{\mcitedefaultmidpunct}
{\mcitedefaultendpunct}{\mcitedefaultseppunct}\relax
\EndOfBibitem
\bibitem[Melo \latin{et~al.}(2024)Melo, T.~Cuairan, Tomassi, Meyer, and
  Quidant]{Melo24}
Melo,~B.; T.~Cuairan,~M.; Tomassi,~G.~F.; Meyer,~N.; Quidant,~R. Vacuum
  levitation and motion control on chip. \emph{Nat. Nanotechnol.}
  \textbf{2024}, 1--7\relax
\mciteBstWouldAddEndPuncttrue
\mciteSetBstMidEndSepPunct{\mcitedefaultmidpunct}
{\mcitedefaultendpunct}{\mcitedefaultseppunct}\relax
\EndOfBibitem
\bibitem[Lepeshov \latin{et~al.}(2023)Lepeshov, Meyer, Maurer, Romero-Isart,
  and Quidant]{Lepeshov23}
Lepeshov,~S.; Meyer,~N.; Maurer,~P.; Romero-Isart,~O.; Quidant,~R. Levitated
  Optomechanics with Meta-Atoms. \emph{Phys. Rev. Lett.} \textbf{2023},
  \emph{130}, 233601\relax
\mciteBstWouldAddEndPuncttrue
\mciteSetBstMidEndSepPunct{\mcitedefaultmidpunct}
{\mcitedefaultendpunct}{\mcitedefaultseppunct}\relax
\EndOfBibitem
\bibitem[Geng \latin{et~al.}(2020)Geng, Peters, Trichet, Malmir, Kolkowski,
  Smith, and Rodriguez]{Geng20}
Geng,~Z.; Peters,~K. J.~H.; Trichet,~A. A.~P.; Malmir,~K.; Kolkowski,~R.;
  Smith,~J.~M.; Rodriguez,~S. R.~K. Universal Scaling in the Dynamic
  Hysteresis, and Non-{M}arkovian Dynamics, of a Tunable Optical Cavity.
  \emph{Phys. Rev. Lett.} \textbf{2020}, \emph{124}, 153603\relax
\mciteBstWouldAddEndPuncttrue
\mciteSetBstMidEndSepPunct{\mcitedefaultmidpunct}
{\mcitedefaultendpunct}{\mcitedefaultseppunct}\relax
\EndOfBibitem
\bibitem[Peters \latin{et~al.}(2021)Peters, Geng, Malmir, Smith, and
  Rodriguez]{Peters21}
Peters,~K. J.~H.; Geng,~Z.; Malmir,~K.; Smith,~J.~M.; Rodriguez,~S. R.~K.
  Extremely Broadband Stochastic Resonance of Light and Enhanced Energy
  Harvesting Enabled by Memory Effects in the Nonlinear Response. \emph{Phys.
  Rev. Lett.} \textbf{2021}, \emph{126}, 213901\relax
\mciteBstWouldAddEndPuncttrue
\mciteSetBstMidEndSepPunct{\mcitedefaultmidpunct}
{\mcitedefaultendpunct}{\mcitedefaultseppunct}\relax
\EndOfBibitem
\bibitem[Rodriguez \latin{et~al.}(2017)Rodriguez, Casteels, Storme,
  Carlon~Zambon, Sagnes, Le~Gratiet, Galopin, Lema\^{\i}tre, Amo, Ciuti, and
  Bloch]{Rodriguez17}
Rodriguez,~S. R.~K.; Casteels,~W.; Storme,~F.; Carlon~Zambon,~N.; Sagnes,~I.;
  Le~Gratiet,~L.; Galopin,~E.; Lema\^{\i}tre,~A.; Amo,~A.; Ciuti,~C.; Bloch,~J.
  Probing a Dissipative Phase Transition via Dynamical Optical Hysteresis.
  \emph{Phys. Rev. Lett.} \textbf{2017}, \emph{118}, 247402\relax
\mciteBstWouldAddEndPuncttrue
\mciteSetBstMidEndSepPunct{\mcitedefaultmidpunct}
{\mcitedefaultendpunct}{\mcitedefaultseppunct}\relax
\EndOfBibitem
\bibitem[Casteels \latin{et~al.}(2017)Casteels, Fazio, and Ciuti]{Casteels17}
Casteels,~W.; Fazio,~R.; Ciuti,~C. Critical dynamical properties of a
  first-order dissipative phase transition. \emph{Phys. Rev. A} \textbf{2017},
  \emph{95}, 012128\relax
\mciteBstWouldAddEndPuncttrue
\mciteSetBstMidEndSepPunct{\mcitedefaultmidpunct}
{\mcitedefaultendpunct}{\mcitedefaultseppunct}\relax
\EndOfBibitem
\bibitem[Tafoya \latin{et~al.}(2019)Tafoya, Large, Liu, Bustamante, and
  Sivak]{Tafoya19}
Tafoya,~S.; Large,~S.~J.; Liu,~S.; Bustamante,~C.; Sivak,~D.~A. Using a
  system’s equilibrium behavior to reduce its energy dissipation in
  nonequilibrium processes. \emph{Proc. Natl. Acad. Sci. U.S.A.} \textbf{2019},
  \emph{116}, 5920--5924\relax
\mciteBstWouldAddEndPuncttrue
\mciteSetBstMidEndSepPunct{\mcitedefaultmidpunct}
{\mcitedefaultendpunct}{\mcitedefaultseppunct}\relax
\EndOfBibitem
\bibitem[Nicholson \latin{et~al.}(2020)Nicholson, Garc{\'\i}a-Pintos, del
  Campo, and Green]{Nicholson20}
Nicholson,~S.~B.; Garc{\'\i}a-Pintos,~L.~P.; del Campo,~A.; Green,~J.~R.
  Time--information uncertainty relations in thermodynamics. \emph{Nat. Phys.}
  \textbf{2020}, \emph{16}, 1211--1215\relax
\mciteBstWouldAddEndPuncttrue
\mciteSetBstMidEndSepPunct{\mcitedefaultmidpunct}
{\mcitedefaultendpunct}{\mcitedefaultseppunct}\relax
\EndOfBibitem
\end{mcitethebibliography}

\providecommand{\latin}[1]{#1}
\makeatletter
\providecommand{\doi}
  {\begingroup\let\do\@makeother\dospecials
  \catcode`\{=1 \catcode`\}=2 \doi@aux}
\providecommand{\doi@aux}[1]{\endgroup\texttt{#1}}
\makeatother
\providecommand*\mcitethebibliography{\thebibliography}
\csname @ifundefined\endcsname{endmcitethebibliography}
  {\let\endmcitethebibliography\endthebibliography}{}

\end{document}